\documentclass[runningheads]{llncs}
\usepackage[T1]{fontenc}
\usepackage{amsmath, amssymb}
\usepackage{hyperref}
\usepackage{graphicx}
\begin{document}
\title{Normative Modeling with Focal Loss and Adversarial Autoencoders for Alzheimer's Disease Diagnosis and Biomarker Identification}
%
\titlerunning{Normative Modeling for AD Diagnosis and Biomarker Identification}

%
\author{Songlin Zhao\inst{1} \and
Rong Zhou\inst{1} \and
Yu Zhang\inst{2}\and
Yong Chen\inst{3} \and
Lifang He\inst{1} 
}
\authorrunning{S. Zhao et al.}
%
\institute{Department of Computer Science and Engineering, Lehigh University,\\Bethlehem, PA, USA\\ \email{lih319@lehigh.edu} \and
Department of Bioengineering, Lehigh University, Bethlehem, PA, USA \and
Department of Biostatistics, Epidemiology and Informatics,\\
University of Pennsylvania, Philadelphia, PA, USA}
\maketitle              
\begin{abstract}

In this paper, we introduce a novel normative modeling approach that incorporates focal loss and adversarial autoencoders (FAAE) for Alzheimer's Disease (AD) diagnosis and biomarker identification. Our method is an end-to-end approach that embeds an adversarial focal loss discriminator within the autoencoder structure, specifically designed to effectively target and capture more complex and challenging cases. We first use the enhanced autoencoder to create a normative model based on data from healthy control (HC) individuals. We then apply this model to estimate total and regional neuroanatomical deviation in AD patients. Through extensive experiments on the OASIS-3 and ADNI datasets, our approach significantly outperforms previous state-of-the-art methods. This advancement not only streamlines the detection process but also provides a greater insight into the biomarker potential for AD. Our code can be found at \url{https://github.com/soz223/FAAE}.

\end{abstract}
\section{Introduction}
\label{sec:intro}

Alzheimer's Disease (AD) is a progressive neurodegenerative disorder characterized by brain dysfunction, presenting significant challenges in both diagnosis and treatment due to individual heterogeneity. Early and accurate detection of AD is crucial for effective patient management and treatment planning. Traditional diagnostic methods primarily rely on clinical assessments and neuroimaging techniques, which can be time-consuming and subjective. To address these challenges, there is an increasing trend toward developing automated, data-driven methods for AD diagnosis and biomarker analysis. 

Normative modeling is a powerful statistical framework for clinical assessments that compares individual deviations against a normative range derived from a healthy control (HC) population \cite{marquand2016understanding}. This method captures variability by comparing with a standard reference model, elucidating disease heterogeneity and uncovering abnormalities. Given the imbalanced nature of medical data \cite{wicaksana2023fca}, where models often bias toward larger groups, normative modeling effectively avoids this by using only one group for training. Recent advancements in deep learning, especially autoencoders (AEs), have further advanced normative modeling. Various methods have been proposed, such as \cite{chamberland2021detecting,chen2018unsupervised,kusner2017grammar,lawry2022conditional,pinaya2019using,pinaya2021using,schlegl2017unsupervised,wang2023normative,wolleb2022diffusion}.

Drawing on insights from previous studies \cite{pinaya2021using,wang2023normative}, we leverage recent developments in the use of adversarial autoencoders (AAEs) for normative modeling. Adversarial learning enhances AEs by aligning the aggregated posterior with the prior, thus minimizing divergence between the model's prior and posterior for improved accuracy. Despite these advances, a notable challenge in existing models is their reduced effectiveness in learning from complex samples, particularly in contexts with uneven data distribution, where some data samples are inherently easier for the model to learn, while others pose significant challenges due to their complexity. This discrepancy often leads to adversarial learning models focusing on simpler patterns that are easier to replicate, neglecting the intricate and complex patterns in the more difficult samples. Additionally, the emphasis on minimizing divergence between the model's prior and posterior can result in reduced sensitivity to these nuanced variations, thereby affecting the model's overall ability to adapt and generalize effectively across varied data samples. This situation highlights the critical need for innovative approaches that can effectively address these specific challenges in normative modeling.

In this paper, we introduce a novel normative modeling approach that leverages focal loss and adversarial autoencoders (FAAE) to enhance the detection of AD. 
By combining these elements, our approach effectively focuses training on challenging cases, preventing easy examples from dominating the training process. We present the results of our extensive testing on the OASIS-3 and ADNI, which are comprehensive and widely used datasets in AD research. Our findings indicate that FAAE-based normative model significantly outperforms previous state-of-the-art methods in AD detection in terms of AUROC (Area Under the Receiver Operating Characteristic Curve score) and sensitivity scores. By analyzing the contrast in deviation plots for AD compared to HC, we can gain a deeper understanding of disease heterogeneity, offering a promising framework for clinical diagnosis and biomarker discovery. This strategic integration bridges the gap between existing methods and the untapped potential of focal loss in normative modeling, setting a new precedent in the field.

\section{Materials and Methods}
\subsection{Dataset Collection and Processing}

In this study, we use fMRI data from Open Access Series of Imaging Studies 3 (OASIS-3) \cite{lamontagne2019oasis} and Alzheimer's Disease Neuroimaging Initiative (ADNI) \cite{mueller2005alzheimer} databases. OASIS-3 comprises a total of 1497 samples with 21 AD samples and 1476 HC samples, and ADNI comprises a total of 579 samples with 141 AD samples and 438 HC samples.

For both OASIS-3 and ADNI datasets, we follow the standard procedures to preprocess each sample using the fMRIPrep pipeline \cite{esteban2019fmriprep}, including intensity nonuniformity correction, skull stripping, spatial normalization, FSL-based segmentation, boundary-based registration, slice-time correction, susceptibility distortion correction, resampling in both original and standard spaces, and motion artifact removal using ICA-AROMA. To handle data collected from multiple periods, we treated the data for each 100-day interval as a sample, assuming no significant change during that period.
Data acquisition during a 6-minute session (164 volumes) employs a 16-channel head coil scanner (TR=2.2 s, TE=27 ms, FOV=240×240 mm, FA=90°). 

In order to generate regional features for each sample, we first average the voxel-level BOLD time series into 100 regions-of-interest (ROIs) for each time point based on the Schaefer-100 parcellation \cite{schaefer2018local}. These averaged time series are then further averaged over time points to create ROI-based input features. Following \cite{pinaya2021using,wang2023normative}, we incorporate key demographic variables including age, gender, and intracranial volume (ICV) as covariates to control their potential impact on the results. Utilizing the same preprocessing steps results in a 22-dimensional covariate vector for each sample. 


\vspace{-5pt}

\subsection{Normative Modeling}
\begin{figure*}[!t]
	\centering
   \includegraphics[width=1\linewidth]{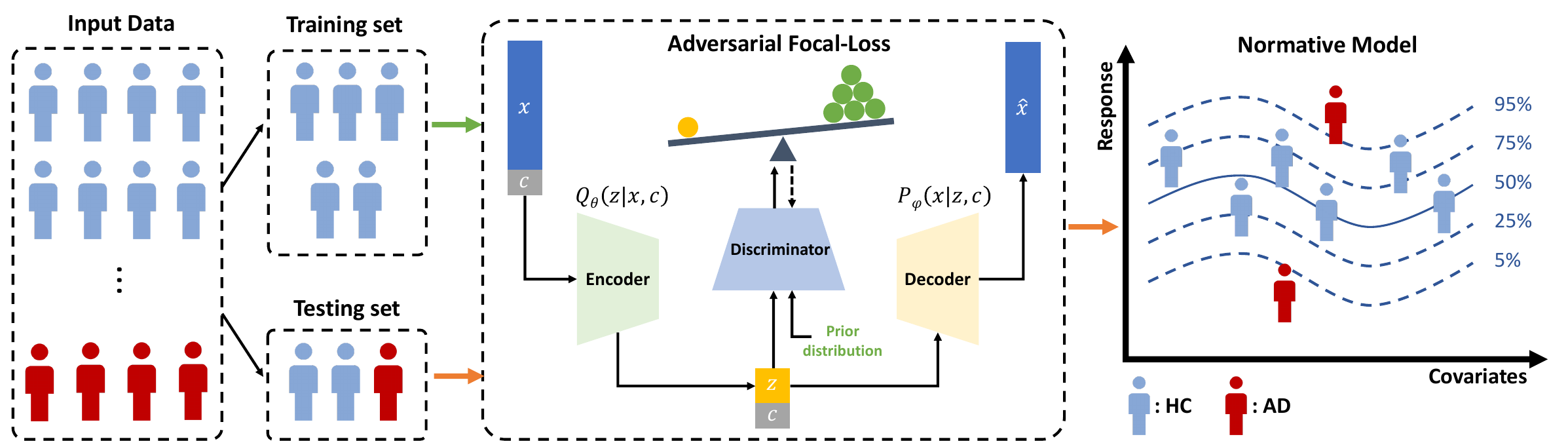}
   \vspace{-5pt}
	\caption{An overview of our proposed FAAE. }
\label{FAAE_overview}
\vspace{-10pt}
\end{figure*}

\textbf{Overview.} Fig.~\ref{FAAE_overview} presents an overview of our proposed FAAE architecture and process for normative modeling. In the training phase, the model is trained only on the HC group, thus constructing a normative range of healthy brain patterns in each brain region. In the testing phase, we calculate a thorough evaluation of each patient's deviation from the established normative range, facilitating precise identification of AD and its associated ROIs.

We use an autoencoder as our model architecture. The backbone module establishes a normative range by training with a dataset of healthy control (HC) individuals. The encoder condenses features into a latent representation, while the decoder reconstructs the input from this latent space, enabling the model to understand healthy brain patterns. An adversarial focal loss discriminator is then integrated to enhance the model's sensitivity in detecting complex AD cases. Below, we detail each module.

\noindent \textbf{Autoencoder.} In this study, we employ the conditional variational autoencoder (CVAE) as the foundational architecture for our normative modeling framework. This type of autoencoder allows us to influence the model's reconstruction using demographic variables such as age, gender, and ICV. Moreover, it is capable of generating a probabilistic latent space representation, which is crucial for effectively capturing the inherent variability and uncertainty prevalent in the data for normative modeling.

Our model architecture consists of three primary components: the encoder, the latent distribution, and the decoder. The encoder compresses high-dimensional data into a compact, low-dimensional latent space. The latent distribution focuses on understanding the data's distribution through its mean ($\boldsymbol{\mu}$) and variance ($\boldsymbol{\sigma}$), which define the probabilistic contours of the latent space. These parameters are essential for generating new data samples, useful for augmenting datasets, especially in cases of rare conditions or imbalanced datasets common in normative modeling. In our study, we use standard random sampling within the latent space to facilitate this generative process as follows:
\begin{equation}
    \mathbf{z} = \boldsymbol{\mu} + \boldsymbol{\sigma} \odot \boldsymbol{\varepsilon},
    \quad  \boldsymbol{\varepsilon} \sim \mathcal{N}(0,1).
    \label{resampling}
\end{equation}
Where $\mathbf{z}$ is the latent representation obtained from the encoder and $\odot$ denotes the element-wise product. This sampling strategy is instrumental in maintaining the balance between data representation accuracy and the flexibility needed for effective normative modeling. Subsequently, the decoder reconstructs the data back to its original high-dimensional form, starting from this probabilistically encoded latent representation. Through this intricate process of encoding, probabilistic modeling in the latent space, and decoding, our model architecture ensures that the most salient features of the data are preserved and accurately represented. The objective function can be formulated as:
\begin{equation}
    L_{\text{CVAE}} = \mathbb{E}[\log P_{\phi}(\mathbf{x}|\mathbf{z}, \mathbf{c})] - KL(Q_{\theta}(\mathbf{z}|\mathbf{x}, \mathbf{c})||P_{\phi}(\mathbf{z}|\mathbf{c})),  
\label{CVAE_loss}
\end{equation}
\noindent where $\mathbf{x}$ represents the input features and $\mathbf{c}$ the confounding variables. The functions $Q_{\theta}(\mathbf{z}|\mathbf{x}, \mathbf{c})$, $P_{\phi}(\mathbf{x}|\mathbf{z}, \mathbf{c})$, and $P_{\phi}(\mathbf{z}|\mathbf{c})$ correspond to the encoder, decoder, and prior distribution, respectively, with $\phi$ and $\theta$ denoting their parameters. The term $\mathbb{E}[\log P_{\phi}(\mathbf{x}|\mathbf{z}, \mathbf{c})]$ measures the reconstruction error, indicating how closely the output matches the input data. The term Kullback-Leibler (KL) divergence, $KL(Q_{\theta}(\mathbf{z}|\mathbf{x}, \mathbf{c})||P_{\phi}(\mathbf{z}|\mathbf{c}))$, assesses the accuracy of the distribution $Q_{\theta}(\mathbf{z}|\mathbf{x}, \mathbf{c})$.

\vspace{3pt}
\noindent \textbf{Adversarial Focal-Loss Discriminator.} Building on the insights from prior studies \cite{pinaya2021using,wang2023normative}, we integrate adversarial learning into the CVAE-based framework described above. This combination enhances the model's reconstruction loss by incorporating the perceptual-level representation capabilities of the discriminator, a key component in adversarial learning.

In adversarial learning, the system has two components: the discriminator and the generator. The discriminator distinguishes between samples from the prior distribution and the CVAE's latent distribution. The generator (also the decoder) produces samples to fool the discriminator. This interaction improves the quality and accuracy of the generated samples. The objective function can be expressed as follows:
\begin{equation}
L_{\text{Adv}} = \mathbb{E}[\log D(\mathbf{z|c})] + \mathbb{E}[\log(1 - D(Q_{\theta}(\mathbf{z}|\mathbf{x}, \mathbf{c})))],
\label{Adv_loss}
\end{equation}

\noindent where $D(\mathbf{z|c})$ is the discriminator, and $Q_{\theta}(\mathbf{x}|\mathbf{z}, \mathbf{c}))$ is the generator, which in this particular case, acts as the encoder role.

Recent research \cite{liu2022adversarial} indicates that discriminators in adversarial learning may sometimes struggle with hard samples, which are particularly challenging for the model to learn. This can impact the model's ability to adapt and generalize effectively across diverse data samples. To address this issue, we introduce focal loss into the adversarial objective function. Focal loss modifies the standard cross-entropy loss by adjusting the weighting of samples within the loss function, offering significant advantages for handling imbalanced datasets and effectively targeting hard-to-learn samples \cite{gao2020incremental}. To elaborate, the focal loss is mathematically defined as follows:
\begin{equation}
    FL(p)=\begin{cases}
-\alpha(1-p)^{\gamma}\log(p),&y=1 \\
-(1-\alpha)p^{\gamma}\log(1-p),&y=0
\label{focal_loss_expand}
\end{cases}
\end{equation}
where $p$ denotes the predicted probability of the true label. The parameter $\alpha$ serves as a scaling factor, while $\gamma$ is employed to amplify the focus on learning from hard samples, where the model is more likely to make errors. 
Leveraging this formulation in Eq.~(\ref{focal_loss_expand}), we adapt the adversarial learning loss to take the following form:
\begin{equation}
\begin{aligned}
    & L_{\text{AdvFL}} = \mathbb{E}\left[-\alpha(1-D(\mathbf{z} | \mathbf{c}))^{\gamma}\log D(\mathbf{z} | \mathbf{c})\right] \\
    & + \mathbb{E}\left[-(1-\alpha)(D(Q_{\theta}(\mathbf{z}|\mathbf{x}, \mathbf{c})))^{\gamma}\log(1-D(Q_{\theta}(\mathbf{z}|\mathbf{x}, \mathbf{c})))\right].
\end{aligned}
\label{loss_focal_adv}
\end{equation}

This extension to the adversarial learning framework integrates the principles of focal loss, optimizing our model's focus on the more challenging samples encountered while training our one-class normative model on the HC group. Specifically, the parameter $\alpha$ plays a key role in balancing the weights of the discriminator and generator, thus minimizing the bias towards prevalent healthy patterns. Meanwhile, $\gamma$ increases the model's sensitivity to subtle variations, a crucial aspect for identifying early-stage or less apparent anomalies. This approach enhances the model's ability to discern nuanced deviations, which is key for effective normative modeling in medical applications.



\noindent \textbf{Final Loss.} 
Combining Eqs.~(\ref{CVAE_loss}) and (\ref{loss_focal_adv}), the total loss for training our FAAE model is expressed as follows:
\begin{equation}
\vspace{-5pt}
L_{\text{FAAE}} = L_\text{CVAE} + L_\text{AdvFL}.
\label{total_loss}
\end{equation}

\noindent \textbf{Deviation Metric.} We employ the standard mean square error (MSE) as a performance function to compute the deviation between the input data and the reconstructed output, defined as: $D_\text{MSE} = \frac{\| \mathbf{x} - \hat{\mathbf{x}} \|_2^2}{n}$,
where $\hat{\mathbf{x}}$ represents the reconstruction of the input data $\mathbf{x}$ as generated by the decoder, and $n$ denotes the dimension of $\mathbf{x}$, which is set to 100 in this study.

\section{Experiments and Results}
\label{experiment_settings}
\textbf{Experimental Settings.} We split the data into a training set, comprising 80\% of the randomly selected HC samples, and a test set, consisting of the remaining HC samples and all AD samples. We follow the same setting as \cite{wang2023normative} to normalize the training and test sets, as well as neural network architectures, and standard parameter settings. To ensure robust results, we employ bootstrap resampling, repeating the process 30 times and reporting the average results. 


\noindent \textbf{Competing Methods.} To evaluate our proposed FAAE, we conduct comparisons with five deep normative modeling methods: vanilla AE \cite{chamberland2021detecting}, VAE \cite{kingma2013auto}, CVAE \cite{sohn2015learning}, ACVAE \cite{wang2023normative}, and AAE \cite{pinaya2020normative}. Each method represents a unique normative modeling approach. For these methods, we utilize their publicly available codes and apply the same parameter settings as our experiments to ensure a fair and consistent comparison. 


\noindent \textbf{Evaluation Metrics.} Measures of performance included the area under the receiver operating characteristic (AUROC), sensitivity, and specificity, which are commonly used in disease diagnosis. A highly sensitive test ensures that patients with the disease are correctly identified, while a highly specific test ensures that patients without the disease are accurately excluded. The AUROC score combines both sensitivity and specificity, providing a single metric that reflects the overall diagnostic performance of a test. A higher AUROC score indicates better discrimination between patients with and without the disease.
\vspace{-8pt}
\begin{table}[]
\renewcommand\arraystretch{1.3}
\setlength\tabcolsep{3pt}
  \centering
  \caption{Testing performance comparison of different models.}
  \vspace{-5pt}
  \scalebox{0.82}{
  \begin{tabular}{c|ccc|ccc}
\hline
                & \multicolumn{3}{c|}{OASIS-3}                                                                                                & \multicolumn{3}{c}{ADNI}                                                                                                      \\ \hline
Methods          & \multicolumn{1}{c|}{AUROC}               & \multicolumn{1}{c|}{Sensitivity}            & \multicolumn{1}{c|}{Specificity} & \multicolumn{1}{c|}{AUROC}                 & \multicolumn{1}{c|}{Sensitivity}             & \multicolumn{1}{c}{Specificity} \\ \hline
AE              & \multicolumn{1}{c|}{58.83 ± 2.21}          & \multicolumn{1}{c|}{55.24 ± 13.67}          & 61.03 ± 11.29                    & \multicolumn{1}{c|}{65.70$\pm$2.50}          & \multicolumn{1}{c|}{66.60$\pm$5.08}          & 66.93$\pm$5.76                  \\ \hline
VAE             & \multicolumn{1}{c|}{61.54 ± 1.82}          & \multicolumn{1}{c|}{65.71 ± 5.55}           & 56.47 ± 5.40                     & \multicolumn{1}{c|}{59.36$\pm$2.61}          & \multicolumn{1}{c|}{55.46$\pm$2.10}          & \textbf{72.39$\pm$3.52}                  \\ \hline
CVAE            & \multicolumn{1}{c|}{62.81 ± 1.26}          & \multicolumn{1}{c|}{68.10 ± 5.65}           & 56.64 ± 3.42                     & \multicolumn{1}{c|}{62.11$\pm$1.17}          & \multicolumn{1}{c|}{56.60$\pm$7.40}          & 72.39$\pm$8.16                  \\ \hline
ACVAE           & \multicolumn{1}{c|}{64.64 ± 2.53}          & \multicolumn{1}{c|}{64.76 ± 17.97}          & 61.46 ± 13.84                    & \multicolumn{1}{c|}{\textbf{67.82}$\pm$\textbf{0.98}}          & \multicolumn{1}{c|}{67.38$\pm$1.74}          & 64.41$\pm$1.51                  \\ \hline
AAE & \multicolumn{1}{c|}{55.94 ± 3.05}          & \multicolumn{1}{c|}{44.76 ± 21.53}          & \textbf{69.15} ± \textbf{20.47}                    & \multicolumn{1}{c|}{64.57$\pm$2.65}          & \multicolumn{1}{c|}{60.43$\pm$10.35}         & \textbf{72.46$\pm$9.65}                  \\ \hline
FAAE            & \multicolumn{1}{c|}{\textbf{68.56 $\pm$ 3.98}} & \multicolumn{1}{c|}{\textbf{70.00 $\pm$ 12.06}} & \textbf{61.76 $\pm$ 11.99}                   & \multicolumn{1}{c|}{\textbf{66.15}$\pm$\textbf{1.17}}          & \multicolumn{1}{c|}{\textbf{72.20$\pm$5.30}} & 60.50$\pm$4.74                  \\ \hline
\end{tabular}
}
\label{Performance_Comparison}
\vspace{-10pt}
\end{table}

\noindent \textbf{Results.} Table \ref{Performance_Comparison} shows the performance of six comparison methods on both OASIS-3 and ADNI datasets. From the results, we can observe that our method significantly outperforms other methods in terms of sensitivity, while also achieving higher or comparable AUROC scores. In most medical applications, high sensitivity is crucial as it minimizes false negatives, thereby reducing the risk of missing disease cases.

Notably, FAAE demonstrates a marked improvement in sensitivity compared to adversarial learning-based methods like ACVAE and AAE. While adversarial learning typically yields higher specificity in AD detection, as evidenced by AAE's top scores, it significantly falls short in sensitivity. This shortfall is likely due to a bias towards easy samples in adversarial learning, which is especially prevalent in datasets with a dominant healthy class. Confirming our hypothesis, FAAE shows enhanced sensitivity, underscoring the role of focal loss in effectively classifying minority cases in imbalanced datasets. Another observation is that although AAE has high specificity, their sensitivity is low. The reason behind this is that AAE applies adversarial learning on an imbalanced dataset, which leads to overfitting.

\begin{figure*}[!t]
	\centering
    \includegraphics[width=1\linewidth]{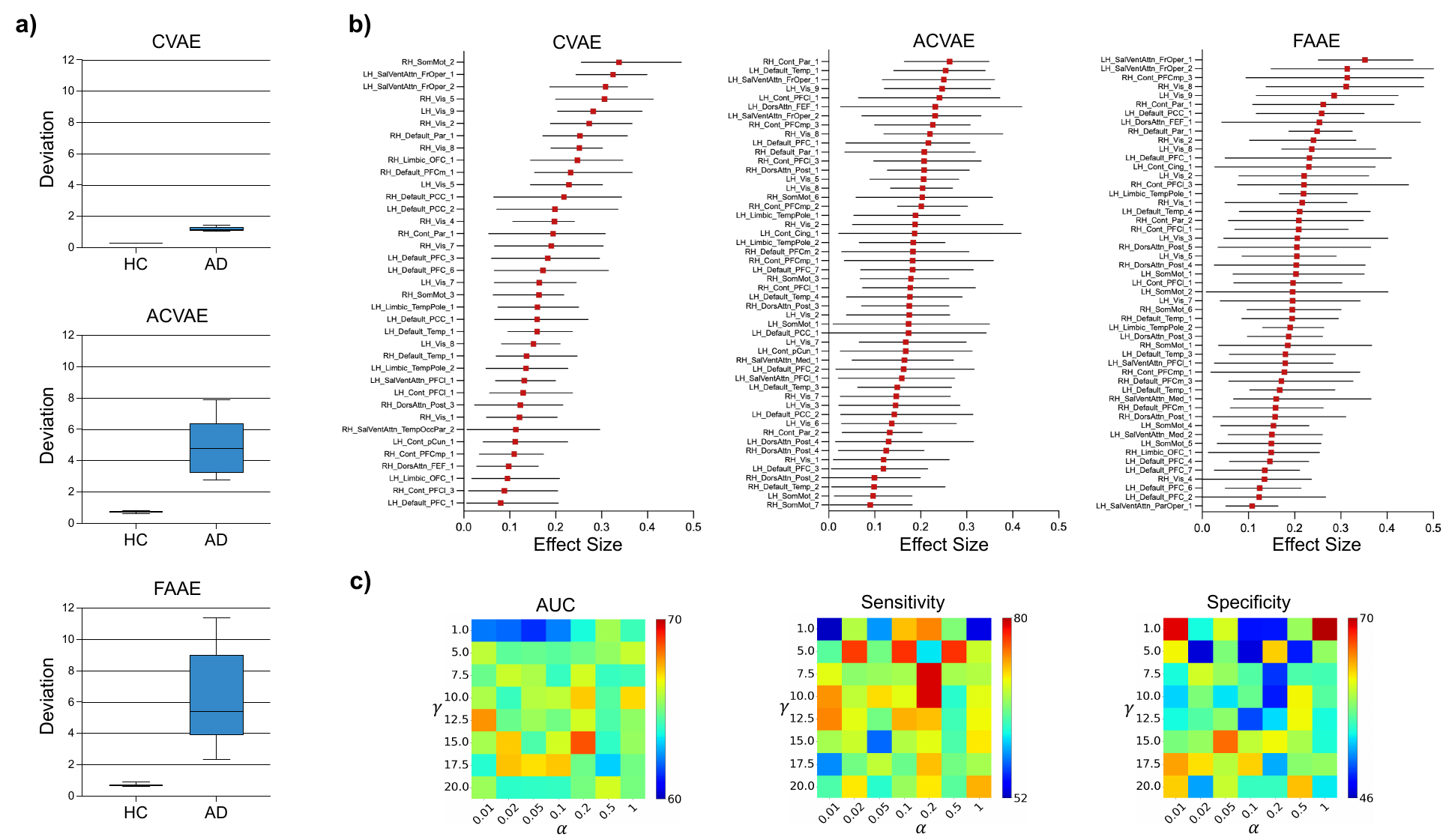}
   \vspace{-18pt}
	\caption{(a)-(b) Observed mean deviation and effect size of HC vs. AD on OASIS-3 for top-3 methods, respectively. (c) Parameter sensitivity on OASIS-3.}
\label{boxlot_effectsize_sensitivityanalysis}
\vspace{-5pt}
\end{figure*}

Moreover, we use OASIS-3 as an example to illustrate our analysis of model performance and regional brain impacts. Fig. \ref{boxlot_effectsize_sensitivityanalysis}(a) shows the deviation boxplots for top-3 methods in the test set, which indicates that FAAE can better distinguish HC and AD. Furthermore, we investigate the impact of different brain regions by computing 95\% confidence intervals for the effect size differences between HC and AD, where an interval not containing 0 indicates a significant difference. Fig.~\ref{boxlot_effectsize_sensitivityanalysis}(b) shows our method identifies more regions with significant effects than baseline models, indicating its superior robustness and sensitivity. 

\begin{figure*}[!t]
	\centering
	  \includegraphics[width=1\linewidth]{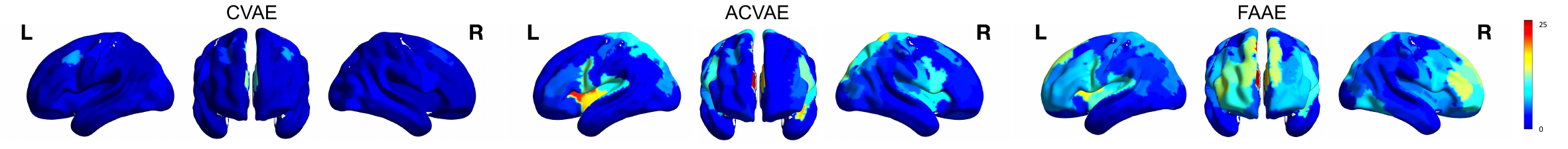}
   \vspace{-10pt}
	\caption{Comparative analysis of average regional-deviation values of AD across top-3 methods on OASIS-3 dataset.}
\label{brainplot_img}
\vspace{-10pt}
\end{figure*}

Additionally, we demonstrate regional variations for the top-3 methods by calculating deviations between expected norms and observed values in AD. Fig. \ref{brainplot_img} shows the results on OASIS-3. Regions with higher deviations indicate a stronger association with AD. Notably, FAAE identifies critical areas like LH\_Default\_PFC\_5 (prefrontal cortex) and RH\_Cont\_PFCmp\_1 (medial posterior prefrontal cortex), consistent with previous studies \cite{jobson2021role,wang2007altered,wang2015differentially}, and uncovers additional AD-related regions such as RH\_Default\_pCunPCC\_2 (precuneus posterior cingulate cortex), RH\_Cont\_PFCl\_2 (lateral prefrontal cortex), and RH\_Cont\_Cing\_1 (cingulate cortex). On the ADNI dataset, FAAE also identifies critical brain regions such as LH\_Default\_PFC\_6 (prefrontal cortex) and RH\_Cont\_PCFmp\_2 (medial posterior prefrontal cortex), aligning with previous studies \cite{grady2003evidence,jobson2021role}. Particularly, FAAE discovers somatosensory dysfunction in RH\_SomMot\_5 (primary somatosensory cortex). These findings, highlighting both known and novel regions linked to AD, could be crucial for identifying potential biomarkers in Alzheimer's research.

\noindent \textbf{Parameter Analysis.} Fig. \ref{boxlot_effectsize_sensitivityanalysis}(c) demonstrates the parameter sensitivity analysis of FAAE with respect to the focal loss parameters $\alpha$ and $\gamma$ on OASIS-3. The results reveal stable AUROC performance across most settings, with a noticeable increase in AUROC values at higher $\gamma$ levels, particularly $\gamma = 15$ and $\gamma = 17.5$ combined with specific $\alpha$ values. Sensitivity to $\alpha$ and $\gamma$ is evident, with higher detection rates of actual AD cases at increased $\gamma$, especially at $\gamma = 15$ and $\alpha = 0.2$. This underscores the importance of carefully balancing $\alpha$ and $\gamma$ to optimize the model for medical diagnostics.

\noindent \textbf{Sample Size Analysis.}
We investigate the effect of varying HC sample sizes in the training set. Table \ref{hc_compare} presents the performance of our FAAE model on the ADNI dataset. All training samples are sampled from the 80\% HC as in the experimental setting. Generally, increasing the number of HC samples in the training data enhanced model performance initially, but this improvement stabilized as the sample size continued to grow. For instance, AUROC improves from 65.98$\pm$2.36 with 200 samples to 70.04$\pm$1.86 with 1000 samples, with no significant improvement beyond that. Specificity follows a similar trend. Sensitivity increases from 65.20$\pm$5.02 with 200 samples to 70.57$\pm$4.06 with 600 samples, then slightly decreases to 68.35$\pm$4.60 at 1400 samples.


\begin{table}[]
\vspace{-15pt}
\renewcommand\arraystretch{1.3}
\setlength\tabcolsep{3pt}
  \centering
  \caption{Performance as a function of the number of training samples.}
  \vspace{-5pt}
  \scalebox{0.82}{
  \begin{tabular}{c|c|c|c|c|c|c|c}
\hline
Metrics     & 200            & 400            & 600            & 800            & 1000           & 1200           & 1400           \\ \hline
AUROC        & 65.98$\pm$2.36 & 67.73$\pm$2.45 & 68.91$\pm$3.12 & 69.05$\pm$2.23 & 70.04$\pm$1.86 & 69.85$\pm$1.32 & 70.01$\pm$1.59 \\ \hline
Sensitivity & 65.20$\pm$5.02 & 65.39$\pm$4.95 & 70.57$\pm$4.06 & 70.71$\pm$5.67 & 68.79$\pm$4.88 & 68.32$\pm$5.92 & 68.35$\pm$4.60 \\ \hline
Specificity & 62.79$\pm$3.66 & 64.84$\pm$3.23 & 62.31$\pm$4.72 & 60.91$\pm$5.92 & 64.82$\pm$3.55 & 64.73$\pm$5.96 & 65.29$\pm$5.05 \\ \hline
\end{tabular}
}
\label{hc_compare}
\vspace{-15pt}
\end{table}

\vspace{-15pt}
\section{Conclusions}
This paper introduces an innovative normative modeling approach for Alzheimer's Disease (AD) diagnosis and biomarker identification. Our method combines an adversarial focal loss discriminator with an autoencoder framework, improving the detection of complex AD cases. By establishing a normative model based on healthy controls, we estimate neuroanatomical deviations in AD patients. Extensive validation on the OASIS-3 and ADNI datasets demonstrates that our approach significantly outperforms existing methods in AD detection, enhancing clinical sensitivity. Future research could expand this work by integrating multimodal data and advanced brain network analysis for a more comprehensive understanding and improved diagnostic precision.


\noindent \textbf{Prospect of Application}: Our FAAE-based normative modeling approach enhances AD diagnosis and biomarker discovery. It aids early detection, personalized treatment, and diagnostic interpretation in clinical settings. In research, it uncovers novel disease mechanisms, improving patient outcomes and advancing the understanding of neurodegenerative diseases.

\begin{credits}
\subsubsection{\ackname} This work is partially supported by the NSF grants (MRI-2215789, IIS-2319451), NIH grants (R21EY034179, R21AG080425, R21MH130956,\\ R01MH129694), and Lehigh's grants under Accelerator (S00010293), CORE (001250), and FIG (FIGAWD35).



\subsubsection{\discintname} The authors have no competing interests to declare that are relevant to the content of this article.
\end{credits}
%
%
%
\bibliographystyle{splncs04}
\bibliography{reference}
%




\end{document}